\documentclass[prd,aps,preprint,nofootinbib]{revtex4}
\usepackage{amsmath,amssymb,graphicx}
\usepackage{color}
\newcommand{\beq}{\begin{equation}}
\newcommand{\eeq}{\end{equation}}
\newcommand{\bea}{\begin{eqnarray}}
\newcommand{\eea}{\end{eqnarray}}
\def\lla{\left\langle}
\def\rra{\right\rangle}

\def\ssc{\scriptscriptstyle}
\def\lsim{\mathrel{\raise.3ex\hbox{$<$\kern-.75em\lower1ex\hbox{$\sim$}}} }
\def\gsim{\mathrel{\raise.3ex\hbox{$>$\kern-.75em\lower1ex\hbox{$\sim$}}} }

\begin{document}

\preprint{{\vbox{\hbox{NCU-HEP-k062} \hbox{Jul 2015}
}}}

\title{A Simple Model of Dynamical Supersymmetry Breaking with the
Generation of Soft Mass(es)}

\author{Yifan Cheng }
\email{yifancheng@cc.ncu.edu.tw}

\affiliation{
Department of Physics,  National Central
University, Chung-Li, Taiwan 32054, \\
and Institute of Physics, Academia Sinica, Taipei,
Taiwan 115 
}

\author{ Yan-Min Dai }

\affiliation{
Department of Physics, National Central
University, Chung-Li, Taiwan 32054 }

\author{ Gaber Faisel }
\email{gfaisel@hep1.phys.ntu.edu.tw}

\affiliation{
Department of Physics,
National Taiwan University, Taipei, Taiwan 10617, \\
and Egyptian Center for Theoretical Physics, Modern University for
Information and Technology, Cairo, Egypt 11211 }

\author{ Otto C. W. Kong }
\email{otto@phy.ncu.edu.tw}

\affiliation{Department of Physics and 
Center for Mathematics and Theoretical Physics, National Central
University, Chung-Li, Tawian, 32054
}


\begin{abstract}
We introduce a simple model of dynamical supersymmetry breaking. It is
like a supersymmetric version of a Nambu--Jona-Lasinio model with a
spin one composite. The simplest version of the model as presented
here has a single chiral superfield (multiplet) with a four-superfield
interaction. The latter has the structure of the square of the
superfield magnitude square. A vacuum condensate of the latter
is illustrated to develop giving rise to supersymmetry breaking
with a soft mass term for the superfield. We report also the effective
theory picture with a real superfield composite, illustrating the matching
effective potential analysis and the vacuum solution conditions for
the components. The nature of its fermionic part as the Goldstone mode
is presented. Phenomenological application to the
supersymmetric standard model is plausible.
\end{abstract}

\maketitle

\section{Introduction}
Our group has recently re-visited supersymmetric versions of the classic
model of dynamical symmetry breaking --- the Nambu--Jona-Lasinio (NJL)
model \cite{NJL}. We have formulated an approach to derive the gap
equation(s) for the supersymmetric mass parameter(s) including
supersymmetry breaking parts \cite{042}. Applying it with a
supergraph calculation, we have succeeded in obtaining the model gap
equations for both the old supersymmetric Nambu--Jona-Lasinio (SNJL)
model \cite{BL,BE} and the new holomorphic version (dubbed HSNJL model)
we proposed as the alternative supersymmetrization \cite{034}. Along
the line of the NJL model idea, the supersymmetric models have dimension
five or six four-superfield interactions which in the regime of large enough
coupling induce the formation of two-superfield condensates that break
some part of the model symmetry and generate masses. In all such cases
soft supersymmetry breaking masses are needed. Exact supersymmetry would
otherwise protect the model against any dynamical symmetry breaking.

Here in this letter, we report on an even more interesting possibility
for the physics of the kind of four-superfield interactions ---
dynamical breaking of supersymmetry itself with the generation of soft
supersymmetry breaking mass(es).

Interesting simple models of spontaneously breaking of supersymmetry are
difficult to find \cite{rev}. A simple model that has the supersymmetry
broken dynamically is even more valuable. Apart from of theoretical
interest, model of soft supersymmetry breaking could be relevant for
TeV scale phenomenology as a background model behind a softly
broken supersymmetric standard model (SSM) \cite{soft}. We have shown that a
HSNJL model is a phenomenological viable version of the SSM with the
Higgs superfields as dynamical composites of quark superfields \cite{034}.
It will be more interesting if the required soft supersymmetry breaking
(squark) masses themselves could be the result of the kind of dynamical
supersymmetry breaking. A model incorporating such a mechanism may be
possible with or even without a single extra superfield beyond that of
the (minimal) SSM.  In contrast, models in the literature usually
require the construction of elaborated supersymmetry breaking and
mediating sectors to accomplish the generation of the soft supersymmetry
breaking terms \cite{soft}. The new model mechanism reported here
hence provides an interesting alternative with plausible
implications for searches at the LHC.

In this letter, we focus on the first step, presenting a prototype
model of such a dynamical supersymmetry breaking with a SNJL type
four-superfield interaction. We skip most of the details of
the calculations here to focus on the key formulational aspects and
the main results. For the skipped details and more general analysis,
please see our companion long paper \cite{063}. A brief discussion
on its possible phenomenological applications will be given at the end.

\section{The model and the soft mass gap equation}
The SNJL model has the four-superfield interaction \cite{BL,BE,042}
\beq \label{snjl}
g^2  \int\!\! d^4 \theta \, \Phi_{+a}^\dagger\Phi_-^{a\dagger} \Phi_+^b\Phi_{-b}
\, (1- \tilde{m}_{\!\ssc C}^2 \theta^2 {\bar \theta}^2 ) \;,
\eeq
in which $a$ and $b$ represent the color indices explicitly shown here.
As the two-superfield condensate
$\lla \left. \Phi_{+a}^\dagger\Phi_-^{a\dagger}\right|_F \rra$
develops, Dirac mass for $\Phi_+^b\Phi_{-b}$ is resulted. Now consider a
similar but supersymmetric interaction with an alternative color index
contraction as given by
\beq \label{new}
g^2  \int\!\! d^4 \theta \, \Phi_{+a}^\dagger \Phi_-^{b\dagger} \Phi_+^a \Phi_{-b}
 \;.
\eeq
One can easily see that if the two-superfield condensate
$\lla \left. \Phi_{+a}^\dagger\Phi_+^{a}\right|_D \rra$ develops, we will
obtain a soft supersymmetry breaking mass for $\Phi_-^{b\dagger} \Phi_{-b}$
(~$ |_D$ denote the component $D$-term, {\it i.e.} the $\theta^2 {\bar \theta}^2$
part, as commonly used in supersymmetric theories). We analyze here the scenario
in the simplest case with only one chiral superfield multiplet, in relation
to the question of dynamical supersymmetry breaking. The very naive looking
model actually gives a highly nontrivial and not very conventional model
Lagrangian in terms of component fields.
\footnote{
In the naive case of really a single superfield, the gap
equation analysis here would correspond to the quenched planar approximation
of QED by Bardeen {\it et.al.} \cite{Bll}, which is commonly believed to give
the correct qualitative result in the kind of dynamical symmetry breaking studies.
Taking $\Phi$ as a fundamental $SO(N)$ multiplet the analysis would have the
usual flavor of an $1/N$ approximation. To have $SU(N)$ `colored-quark'
multiplet, we would have to restrict to the special, but most interesting
case of $m=0$.  One may also then consider the two-superfield case with
a Dirac mass term $m_{\!\ssc D} \Phi_+ \Phi_-$ instead, to retrieve
an $1/N$ approximation structure. Some discussion of the issue is
available in Ref.\cite{050}. Note that a nonzero $m$ or $m_{\!\ssc D}$
is really not necessary for our key result here.
}

Suppressing the color index, the simple model is given with
kinetic term, mass term, and the dimension six interaction, {\it i.e.} as
\beq \label{L}
{\cal{L}}=
\int  d^4 \theta \, \left[
\Phi^\dagger \Phi  \, +\frac{m}{2}   \Phi \Phi \delta^2(\bar{\theta})
+ \frac{m^*}{2} \Phi^{\dagger} \Phi^{\dagger} \delta^2(\theta)
-  \frac{g^2 }{2}   \left( \Phi^\dagger \Phi \right)^2 \right] \;.
\eeq
\begin{figure}[!t]
\begin{center}
\includegraphics[height=3.0cm,width=12.0cm]{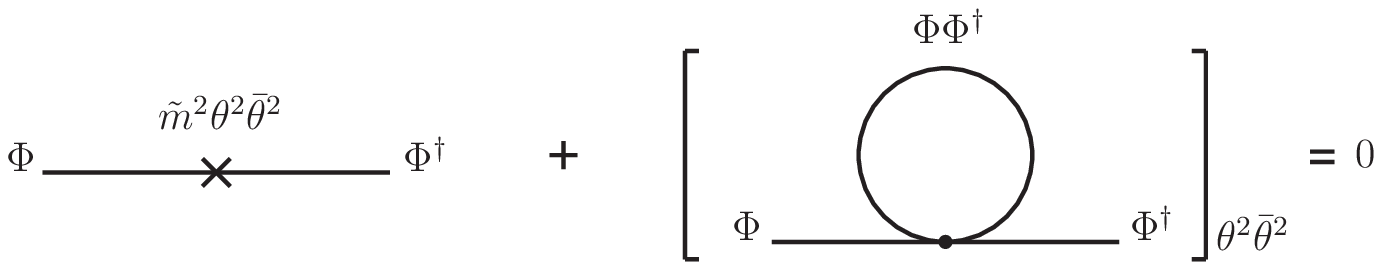}
\end{center}
\caption{\small  The soft mass gap equation. Only the $\theta^2\bar\theta^2$
component of the superfield amplitude for the second part is to be taken.
The other way, to look at it is to see the first part as a $\tilde{m}^2$
insertion for $AA^*$ and the second part as the $\Phi$-loop amplitude for $AA^*$.}
\vspace*{.1in}
\hrule
\end{figure}
It is supersymmetric. Note that the four-superfield interaction is
written with a sign opposite to that of the old SNJL model, the reason
behind which will surface below. Otherwise, one may also look at much of
the analysis without restricting to a positive $g^2$.
The first step of the self-consistent Hartree
approximation is to add the interested soft mass term
$-\int d^4 \theta \Phi^\dagger \Phi \tilde{m}^2 \theta^2 \bar{\theta}^2$
to the free field part and re-subtract it as a mass-insertion type
interaction. The formal gap equation can be illustrated
diagrammatically as in Fig.1, with the analytical expression given by
\beq
\tilde{m}^2 = \left.\Sigma_{\tilde{m}}^{\tiny(loop)}(p)\right|_{\mbox{\tiny on-shell}}
\equiv  -\left.
 \Sigma^{({\mbox \tiny loop})}_{\Phi\Phi^\dagger}(p;\theta^2, \bar{\theta}^2)
 \right|_{ \theta^2 \bar{\theta}^2, \mbox{\tiny on-shell}}  \;.
\eeq
where we have, in accordance with the formulation discussed in Ref.\cite{042},
introduced the superfield two-point proper vertex
$\Sigma_{\Phi\Phi^\dagger}(p;\theta^2, \bar{\theta}^2)$ with both
supersymmetric and supersymmetry breaking parts. The full superfield
theoretical description of the model gap equation will be addressed in
Sec.~IV below. One may go to the component field Lagrangian to evaluate
$\left.\Sigma_{\tilde{m}}^{\tiny(loop)}(p)\right|_{\mbox{\tiny on-shell}}$.
However, the proper self-energy of the scalar $A$ has also a part that
gives a wavefunction renormalization. The latter is supersymmetric,
showing up also for the proper self-energy of the fermion and the
auxiliary components. That is exactly the supersymmetric part of
$- \left.\Sigma^{({\mbox \tiny loop})}_{\Phi\Phi^\dagger}(p;\theta^2, \bar{\theta}^2)
\right|_{\mbox{\tiny on-shell}}$.
The correct gap equation result for $\tilde{m}^2$ can be obtained
after careful treatment of the wavefunction renormalization. A direct
supergraph evaluation of
$\left. \Sigma^{({\mbox \tiny loop})}_{\Phi\Phi^\dagger}(p;\theta^2, \bar{\theta}^2)
 \right|_{\mbox{\tiny on-shell}}$, can then be performed.
 The gap equation result for the soft mass reads
\begin{figure}[!t]
\begin{center}
\includegraphics[height=7.0cm,width=8.0cm]{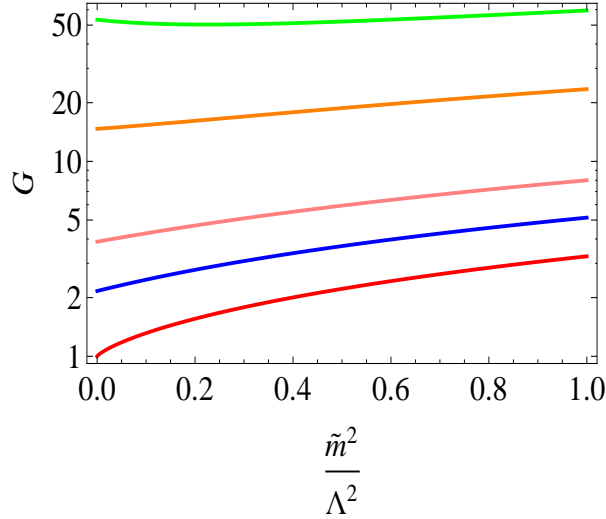}
\end{center}
\caption{\small Numerical plot  of nontrivial solutions to the soft mass
gap equation. Coupling parameter $G$ is plotted against the normalized
soft mass parameter $s\;\left(=\frac{\tilde{m}^2}{\Lambda^2}\right)$
for  $t\;\left(=\frac{{|m|}^2}{\Lambda^2}\right)$ values of $0$ (red),
$0.1$ (blue), $0.2$ (pink), $0.4$ (orange), $0.5$ (green), from the
lowest to the highest curves, respectively.
Notice that the critical coupling increases from $G=1$ for nonzero
values of the input supersymmetric mass $m$.}
\vspace*{.1in}
\hrule
\end{figure}
\beq\label{gap}
\tilde{m}^2 = {g^2} I(m^2, \tilde{m}^2, \Lambda^2) \;,
\eeq
where
\bea
 I(m^2, \tilde{m}^2, \Lambda^2) &=&
\int^{\ssc E} \frac{d^4k}{(2\pi)^4}
\frac{\tilde{m}^2(k^2-m^2)}
{(k^2+m^2)(k^2+m^2+\tilde{m}^2)}
\nonumber \\
&=&
\frac{1}{16\pi^2}
\left[
\Lambda^2 \tilde{m}^2 +2 m^4 \ln{\frac{\Lambda^2 + m^2}{m^2}}
 \right. \nonumber\\
&& \left. \hspace*{.7in}
- (2m^4+3m^2\tilde{m}^2+\tilde{m}^4)
\ln{\frac{(\Lambda^2+m^2+\tilde{m}^2)}{(m^2+\tilde{m}^2)}}
\right] \;,
\eea
in which the (Euclidean) momentum loop integral is evaluated with the
cut-off $\Lambda$. To check for nontrivial solution, we re-write the
equation in dimensionless variables normalized to $\Lambda$
\beq
 \frac{1}{G} = 1+ \frac{2t^2}{s} \ln\left[1+\frac{1}{t}\right]
-\left(\frac{2t^2}{s}+3t+s\right) \ln\left[1+\frac{1}{t+s}\right]
\eeq
where $G=\frac{g^2  \Lambda^2}{16\pi^2}$,
$s=\frac{\tilde{m}^2}{\Lambda^2}$,
and $t=\frac{m^2}{\Lambda^2}$. Nontrivial solution for the soft mass
($0<s<1$) can be resulted for a large enough coupling, as illustrated
numerically in Fig.~2.

It is interesting to note that at the $m=0$ limit, the gap equation
reduces to the condition that the momentum integral of the Feynman
propagator $\Delta_{\!\ssc F}(k^2,\tilde{m}^2)$ equals $\frac{1}{g^2}$ or
\bea
g^2 \int^{\ssc E} {\Delta_{\!\ssc F}}(k^2,\tilde{m}^2) = 1 \;,
\label{mogap}
\eea
which is the same as the basic NJL model one except with the soft
mass $\tilde{m}^2$ replacing the (Dirac) fermionic mass (see for example
Ref.\cite{BE}) if we take $\frac{g^2}{2}$ as the four-fermion coupling
in the model. This is the $t=0$ case, with critical coupling $G=1$.
Nontrivial solution requires bigger and bigger critical coupling
as the $m$ increases, and become impossible beyond a limit.
Obviously, a $N$ factor is to be multiplied to $g^2$ in the gap
equation if $\Phi$ is a $N$-multiplet, or $G$
taken as $\frac{Ng^2  \Lambda^2}{16\pi^2}$.

\section{Component Field and Effective Theory Picture}
Let us take a look at the component field picture of the model, expanding
$\Phi$ as $A+\sqrt{2}\theta \psi +\theta^2 F$. For simplicity, we drop
any reference of a nontrivial `color' factor in the analysis below and
pretend that $\Phi$ is just a single superfield. The Lagrangian is given by
\bea  {\cal L} &=&
 i\partial_{\mu}\bar{\psi}\bar\sigma^{\mu}\psi
  + \partial^{\mu}A^*\partial_{\mu}A  +  F^*F \,
+ \left(m AF  - \frac{m}{2}\psi\psi +  h.c. \right)
- \frac{g^2}{2} | 2 F A -\psi \psi |^2
\nonumber\\&&
 + 2 g^2 A^*A\, \partial^{\mu}A^* \partial_{\mu}A
 - 2 g^2  i\partial_{\mu}\bar{\psi}\bar\sigma^{\mu}\psi   A^*A
 - 2 g^2  i \bar{\psi} \bar\sigma^{\mu} \psi A \partial_{\mu}A^* \,.
\eea
Notice that the model has a $U(1)_{\ssc R}$ symmetry under which
$A$ and $F$ have charge $+1$ and $-1$. If $m=0$, there will also be
a $U(1)$ $\Phi$-number symmetry.
From the equation of motion for the auxiliary field $F^*$, we have
\beq \label{F}
F = - \frac{ (m^*  + g^2 \psi\psi)  A^*}{1-2g^2 |A|^2 }\;.
\eeq
The somewhat complicated fractional form of $F$ means that the component field
Lagrangian with $F$ eliminated would have less than conventional interaction
terms. Naively, the scalar potential is given by
\beq
-V_s = F^*F - 2g^2 A^*A F^*F + m A F + m^* A^* F^* \;.
\eeq
Eliminating $F$ gives, however,
\beq
V_s= \frac{(|m|^2 - g^4 \psi\psi\bar{\psi}\bar{\psi}) |A|^2}{1-2g^2 |A|^2 }\;
\eeq
which formally no longer involves only the scalar. It is suggestive of a
bifermion condensate which fits in the general picture of the NJL setting.
It is interesting to note that for  $m=0$ the model actually has no pure scalar
part in $V_s$, for any coupling $g^2$. On the other hand, if one neglects
the fermion field part in the above, the potential looks simple enough,
$V_s= \frac{|m|^2 |A|^2}{1-2g^2 |A|^2 }$, with a supersymmetric minimum at
zero $A$. For positive $g^2$, however, it blows up at  $|A|=1/\sqrt{2}g$. For
a perturbative coupling, one expect $1/g$ bigger than the model
cut-off scale $\Lambda$, hence the potential is well behaved within
the cut-off. With strong coupling $g^2$, one cannot be so comfortable.
In fact, for $|A|>1/\sqrt{2}g$, the potential goes negative, contradicting our
expectation for a supersymmetric model. The analysis so far suggests
compatibility with the superfield gap equation results. Let us go on
with the introduction of the effective theory having the composite.

Assuming we do have the composite formation as indicated by the superfield
gap equation, the model can also be appreciated with the following effective
theory picture, again along the line of NJL-type models. We use parameters
$m_o$ and $g_o$ in the place of $m$ and $g$ in the original Lagrangian to
which we add
\beq
\mathcal{L}_s =  \int d^{4} \,\theta \frac{1}{2}(\mu U + g_o \bar{\Phi} \Phi )^2 \;,
\eeq
where $U$ is an `auxiliary' real superfield and mass parameter $\mu$ taken as
real and positive (for $g_o^2>0$).
The equation of motion for $U$, from the full
Lagrangian $\mathcal{L} +\mathcal{L}_s$ gives
\beq
U = - \frac{g_o}{\mu} \bar{\Phi} \Phi \;,
\eeq
showing it as a superfield composite of $\bar{\Phi}$ and $\Phi$.
The condition says the model with $\mathcal{L} +\mathcal{L}_s$ is
equivalent to that of $\mathcal{L}$ alone. Expanding the term in
$\mathcal{L}_s$, we have a cancellation of the dimension six interaction
in the full Lagrangian, giving it as
\beq
\mathcal{L}_{ef\!f}\equiv \mathcal{L} +\mathcal{L}_s
 = \int d^{4}\theta \left[ \bar{\Phi}  \Phi
 + \frac{\mu^2}{2} U^{2} + \mu g_o U \bar{\Phi}\Phi
+\frac{m_o}{2} \Phi^2 \delta(\bar{\theta})
+ \frac{m_o^*}{2}  \bar{\Phi}^2 \delta(\theta) \right] \;.
\label{Le}
\eeq
Obviously, if $\left. U \right|_D$ develops a vacuum expectation value
(VEV), supersymmetry is broken spontaneously and the superfield $\Phi$
gains a soft supersymmetry breaking mass of $-\mu g_o \lla\left. U \right|_D\rra$.
The above looks very much like the standard features of NJL-type model.
Notice that while $U$ does contain a vector component, its couplings
differ from that of the usually studied `vector superfield' which is
a gauge field supermultiplet. That is in addition to having $\mu$ as like
a supersymmetric mass for $U$, which can be compatible only with a
broken gauge symmetry. As such, model with superfield $U$ is not
usually discussed. The superfield can be seen as two parts, as
illustrated by the following component expansion,
\begin{eqnarray}U(x,\theta,\bar\theta ) &=&  \frac{C(x)}{\mu}
+\sqrt{2}\theta \frac{\chi(x)}{\mu} + \sqrt{2} \bar\theta \frac{\bar\chi(x)}{\mu}
+\theta \theta \frac{N(x)}{\mu} +\bar\theta \bar\theta \frac{{N^*}(x)}{\mu}
\nonumber \\  && \;\;\;
+\sqrt{2}\theta \sigma^{\mu}\bar \theta v_{\mu}(x)
+\sqrt{2}\theta \theta \bar\theta  \bar\lambda(x)
+\sqrt{2}\bar\theta \bar\theta \theta   \lambda(x)
+\theta \theta \bar\theta \bar\theta D(x)\; ,
\end{eqnarray}
where the components $C$, $\chi$, and $N$ is the first part
which has the content of like a chiral superfield with however $C$
being real. The $\mu$ factor is put to set the mass dimensions
right. The rest is like the content of a superfield for the usual
gauge field supermultiplet, with $D$ and $v_\mu$ real.  Note that
$N$, $\chi$ and $\lambda$ carry $U(1)_{\ssc R}$ charges -2, -1 and
+1, respectively. The effective Lagrangian in component form is
given by
\bea
{\cal L}_{ef\!f}
&=&
(1+g_o C)\left[
          A^*\Box A
         +i(\partial_\mu\bar{\psi}) \bar{\sigma}^\mu\psi
         + F^* F   \right]
+\frac{m_o}{2}   \left( 2 A F  -\psi\psi  \right)
+\frac{m^*_o}{2}  \left(
   2 A^* F^*  -\bar{\psi}\bar{\psi} \right)
\nonumber\\  &&
+ {\mu}CD  -{\mu}\chi\lambda -{\mu}\bar{\chi}\bar\lambda
+NN^* -\frac{\mu^2}{2}v^{\nu}v_{\nu}
-\mu g_o\psi \lambda A^* - \mu g_o\bar{\psi}\bar{\lambda} A  + \mu g_o D   A^* A
\nonumber\\  &&
-i\frac{g_o}{2}\bar{\psi}\bar{\sigma}^\mu\chi \partial_\mu A
+i\frac{g_o}{2}(\partial_\mu\bar{\psi})\bar{\sigma}^\mu\chi A
-g_o\chi\psi F^* +g_o N A F^*
\nonumber\\  &&
+i\frac{g_o}{2}\bar{\chi}   \bar{\sigma}^\mu\psi  \partial_\mu A^*
-i\frac{g_o}{2} A^*   \bar{\chi}\bar{\sigma}^\mu \partial_\mu\psi
-g_o\bar{\chi}\bar{\psi}F +g_o N^* A^* F
%
%
\nonumber\\  &&
- \frac{\mu g_o}{\sqrt{2}}\eta^{\mu\nu}v_\mu iA^*\partial_\nu A
  + \frac{\mu g_o}{\sqrt{2}}\eta^{\mu\nu}v_\mu i(\partial_\nu A^*)A
  -\frac{\mu g_o}{\sqrt{2}}\eta^{\mu\nu}v_\mu \bar{\psi}\bar{\sigma}_\nu\psi
\;. \label{Lec}
\eea
Notice that $F$, $N$, and $D$ are usual auxiliary components.

In accordance with the `quark-loop' approximation in the (standard) NJL
gap equation analysis and our particular supergraph calculation scheme
above in particular, we consider plausible nontrivial vacuum solution
with nonzero vacuum expectation values (VEVs) for the composite scalars
$C$, $N$, and $D$. An effective potential analysis based on the Weinberg
tadpole method \cite{W,M} is performed with the effective Lagrangian in
component form. Vanishing tadpole conditions can be obtained for the scalar
potential $V(C,N,D)$ up to one-loop level. Notice that $\lla C \rra$ here
corresponds to a superfield wavefunction renormalization term  $Z=1+g_o \lla C \rra$
for $\Phi$ or its components. It is the supersymmetric part of
 $\left.\Sigma^{({\mbox \tiny loop})}_{\Phi\Phi^\dagger}(p;\theta^2, \bar{\theta}^2)
 \right|_{\mbox{\tiny on-shell}} $, an unavoidable part of the one-loop
supergraph in our gap equation calculation for the original Lagrangian
discussion in the previous section. One can see the solution equation
for the $C$-tadpole under the consistent assumption of  $\lla N \rra =0$
gives exactly the $\tilde{m}^2$ gap equation of Eq.(\ref{gap})
for the here renormalized soft mass parameter in terms of the renormalized
mass $m$ and coupling $g$. The treatment in the previous section is
essentially non-perturbative. It is natural that superfield $\Phi$ and
the parameters there correspond to the renormalized quantities in the
present perturbative treatment. The feature will be illustrated more
clearly in the next section. We skip all details here. Interested
readers are refer to the accompanying paper \cite{063} in which the general
case of possible nonzero $\lla N \rra$ will be presented.

\section{Full Superfield Picture}
We present here our formulation based on the advocated strategy of
putting superfield functionals as taking values like constant supefields
admitting supersymmetric breaking parts \cite{042}. The first step
is to add to and subtract from the Lagrangian a term with a full
superfield parameter containing the soft mass $\tilde{m}^2_o$.
We introduce the generic form of the superfield parameter
\beq
{\mathcal{Y}}= y-\tilde{\eta}_o\theta^2
-\tilde{\eta}^*_o \bar{\theta}^2-\tilde{m}^2_o\theta^2 \bar{\theta}^2 \;.
\eeq
For simplicity, we assume $\tilde{\eta}_o=0$ and drop it from further
consideration. The Lagrangian is split as
${\mathcal{L}}={\mathcal{L}}_o +{\mathcal{L}}_{int}$ where
$
{\mathcal{L}}_o = \int  \bar{\Phi} \Phi (1+{\mathcal{Y}})
+\frac{m_o}{2}  \Phi^2 \delta^2(\bar{\theta})
+ \frac{m_o^*}{2}  \bar{\Phi}^2 \delta^2({\theta})
$ 
and
$
{\mathcal{L}}_{int} =   \int  - {\mathcal{Y}}  \bar{\Phi} \Phi
-\frac{g^2_o}{2}     \bar{\Phi} \Phi \bar{\Phi} \Phi  
$, 
in which we have hidden the $d^4\theta$ and use $m_o$ in place of $m$ and for
${g^2_o}$ in place of $g^2$. Obviously, a nonzero $y$ contributes to wavefunction
renormalization $\Phi_{\!\ssc R}\equiv \sqrt{Z} \Phi=\sqrt{1+y}\,\Phi$.
The quantum effective action is given by
\bea
\Gamma &=& \int  \bar{\Phi}_{\!\ssc R} \Phi_{\!\ssc R}
(1-\tilde{m}^2 \theta^2 \bar{\theta}^2)
+\frac{m}{2}  \Phi^2_{\!\ssc R} \delta^2(\bar{\theta})
+ \frac{m^*}{2}  \bar{\Phi}^2_{\!\ssc R} \delta^2({\theta})
\nonumber \\ && \hspace*{.2in}
- \frac{\mathcal{Y}}{Z}     \bar{\Phi}_{\!\ssc R} \Phi_{\!\ssc R}
-\frac{g^2}{2} \bar{\Phi}_{\!\ssc R} \Phi_{\!\ssc R} \bar{\Phi}_{\!\ssc R} \Phi_{\!\ssc R}
+\Sigma_{\Phi_{\!\ssc R}\Phi_{\!\ssc R}^\dagger} \bar{\Phi}_{\!\ssc R} \Phi_{\!\ssc R}
 + \cdots  \;,
\eea
with now renormalized $m$ and $g^2$.
The superfield gap equation under the NJL framework is given by
\bea
- {\mathcal{Y}}_{\!\ssc R} + \left.
  \Sigma_{\Phi_{\!\ssc R}\Phi_{\!\ssc R}^\dagger}^{\tiny(loop)}(p; \theta^2 \bar{\theta}^2)
     \right|_{\mbox{\tiny on-shell}}   =0 \;,
\label{sgap}
\eea
where ${\mathcal{Y}}_{\!\ssc R} \equiv \frac{\mathcal{Y}}{Z}$.
We introduce the expansion
$\Sigma_{\Phi_{\!\ssc R}\Phi_{\!\ssc R}^\dagger}(p; \theta^2 \bar{\theta}^2)
= \Sigma_{r}(p)-  \Sigma_{\tilde{\eta}}(p)  \theta^2
 -  \bar{\Sigma}_{\tilde{\eta}^*}(p)  \bar{\theta}^2
  - \Sigma_{\tilde{m}}(p)   \theta^2 \bar{\theta}^2$
from which we can obtain the gap equation for $\tilde{m}^2$ as Eq.(\ref{gap})
above. Notice that the standard gap equation picture has to be interpreted
here in terms of renormalized superfield and couplings, which is to be
expected in the presence of nonzero wavefunction renormalization. The other
parts of the superfield gap equation read
$\left.\Sigma_{\tilde{\eta}}^{\tiny(loop)}(p)\right|_{\mbox{\tiny on-shell}}=0$
(or rather $=\tilde{\eta}$)  and
$\left.\Sigma_{r}^{\tiny(loop)}(p)\right|_{\mbox{\tiny on-shell}}=\frac{y}{1+y}$.

It is interesting to see that the effective potential analysis for (the
components of) the composite superfield $U$ can be shown directly to be
equivalent to the superfield gap equation. Potential minimum condition
is given by
\bea &&
\mu^2 \lla U \rra + U_{tadpole} =0
 \qquad    \Longrightarrow \qquad
\mu g \lla U \rra =  - g^2 I^{\tiny(loop)}_{\Phi_{\!\ssc R}\Phi_{\!\ssc R}^\dagger}
\eea
where $I^{\tiny(loop)}_{\Phi_{\!\ssc R}\Phi_{\!\ssc R}^\dagger}$ is the momentum
integral of the ${\Phi_{\!\ssc R}\Phi_{\!\ssc R}^\dagger}$ propagator loop.
Note that from the original Lagrangian with two-superfield composite assumed,
we can obtain
$-g^2 \lla \left( \Phi_{\!\ssc R}\Phi_{\!\ssc R}^\dagger \right) \rra
 ={{\mathcal{Y}}_{\!\ssc R}}$,
which is equivalent to $\mu g \lla U \rra = {{\mathcal{Y}}_{\!\ssc R}}
= \left. \Sigma_{\Phi_{\!\ssc R}\Phi_{\!\ssc R}^\dagger}^{\tiny(loop)}(p; \theta^2 \bar{\theta}^2)  \right|_{\mbox{\tiny on-shell}}
 =  - g^2 I^{\tiny(loop)}_{\Phi_{\!\ssc R}\Phi_{\!\ssc R}^\dagger}$.
The same loop integral is of course involved in both the gap equation
picture and the effective potential analysis. The results here are in direct
matching with the correspondent discussion for the NJL case presented in
Ref.\cite{BE}, though for a superfield theory instead. The $\tilde{\eta}$
parameter is to be matched to the $\lla N \rra$ in $\lla U \rra$ and
 $\lla N \rra=0$ can be shown to be a consistent solution in the effective
potential analysis.

\section{the Goldstino}
With the supersymmetry breaking, we expect to have a Goldstino. The required
analysis is the `quark-loop' contribution of the two-point function for
the composite superfield $U$ in addition to the tree-level mass term. The loop
contribution also generates a kinetic term to turn $U$ into a dynamic one.
Here in this letter, however, we are contented with a minimal demonstration
for a special case. We illustrate here the presence of the Goldstone mode for
the special but most interesting case of $m=0$. More details will be left
to Ref.\cite{063}. We use component field calculation with renormalized
$\Phi$ and coupling $g$.

There are two fermionic components of $U$, the $\chi$ and $\lambda$, with the
tree-level Dirac mass term $\mu \chi\lambda$. For the $\lla N \rra=0$, the
$U(1)_{\!\ssc R}$ symmetry is maintained, which protects against any $\chi\chi$
or $\lambda\lambda$ (Majorana) mass term. For $m=0$ then, there is only one diagram
contributing to $\lambda\chi$ mass (see Fig.~3). The diagram has a propagator
for the scalar $A$ component and one for the fermion $\psi$ component
(of the renormalized $\Phi$). The mass produced has a magnitude given by
the momentum integral of the two propagators. The diagram has two copies,
one has an $A$-momentum, the other a $\bar{\psi}$-momentum dependence at the
$\chi$-vertex, with equal and opposite coupling [{\it cf.} Eq.({\ref{Lec})].
At zero external momentum, the two copies give equal contributions which
add up to a mass term of value to be given by
$-2\cdot g \cdot \frac{\mu g}{2} \int^{\ssc E}
    \Delta_{\!\ssc F} (k^2,\tilde{m}^2)$, where
$\Delta_{\!\ssc F} (k^2,\tilde{m}^2)$ is just the $A$-propagator; the
$\psi$-propagator gives only momentum factor(s) that cancels those from
the $\chi$-vertex. From the gap equation as given in Eq.(\ref{mogap})
one can see that the loop generated mass is exactly $-\mu$, which is
equal and opposite to the tree-level mass. So, as the loop tadpoles
cancel the tree-level ones generating the symmetry breaking solution,
an effectively local term from the loop generated two-field proper vertex
cancel the tree-level mass term. The latter is again a generic feature of
NJL models, which should work for the full superfield $U$ in our case. For
the fermionic components, there is no other piece of contribution to
$\chi$ or $\lambda$ masses, hence they are the Goldstino(s). Note that there
are wavefunction renormalization terms for  $\chi$ and $\lambda$ which make
them dynamic. The wavefunction renormalizations do not affect our mass
discussion here. The mass for the un-renormalized   $\chi$-$\lambda$
sector is zero. The renormalized mass would remain zero.

\begin{figure}[!t]
\begin{center}
\includegraphics[height=4.0cm,width=6.0cm]{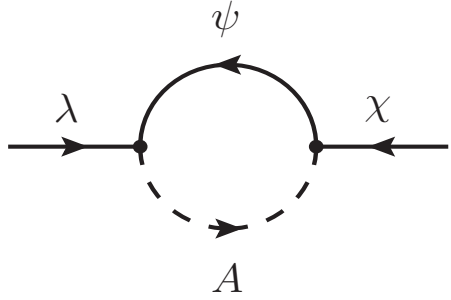}
\end{center}
\caption{\small  Diagram for $\lambda\chi$ mass.}
\vspace*{.1in}
\hrule
\end{figure}

\section{Remarks and Conclusions}
From our superfield model Lagrangian of Eq.(\ref{L}), the interaction term
$\frac{-g^2}{2} \left(\Phi^\dagger\Phi\right) \left(\Phi^\dagger\Phi\right)$
contains component field parts in the form of kinetic terms multiplied to
the scalar field product $A^*A$, which may look very un-conventional.
However, superfields are scalars on the superspace. The interaction is
like a superspace version of four-scalar interaction. It is of dimension six,
but quite natural to be considered in models with a cut-off.
As a term in the superfield Kahl\"er potential,
the interaction term does not look any stranger than the term
$\frac{g^2}{2} \left(\Phi^\dagger\Phi^\dagger\right) \left(\Phi\Phi\right)$,
studied before in the first supersymmetric version of the NJL model \cite{BL,BE},
which contains the usual NJL four-fermion interaction.
The difference is only in the (color) index contraction. On the other hand,
in the literature on the original of soft supersymmetry breaking masses
for the SSM, similar interaction of the form
$-g^2_s \Phi_s^\dagger\Phi_s \Phi^\dagger\Phi$,
where $\Phi_s$ is the so-called spurion superfield bearing a supersymmetry
breaking VEV to communicate supersymmetry breaking to the SSM superfields $\Phi$
has been discussed a lot. Our model can be considered like a similar scenario,
only with the $\lla\left. \Phi_s^\dagger\Phi_s \right|_D\rra$ coming from
the dynamically induced two-superfield condensate instead of individual
$\lla\left.\Phi_s \right|_F\rra$. But then the kind of $g^2_s$ term
in itself can be the very source of supersymmetry breaking and no extra
supersymmetry breaking sector modeling is needed. That should
be a very interesting scenario for phenomenological model building.
It may even be possible to have a kind of simpler models with the role of
$\Phi_s$ played by one of the SSM matter superfields themselves. As
an initiate nonzero supersymmetric mass for $\Phi_s$ is not necessary
for the supersymmetry breaking mechanism to work, the chiral nature
of the SSM quark superfields does not present a problem. In fact,
the $m=0$ is definitely more interesting as it generates mass(es)
from no imput mass scale, like the basic NJL model.

Together with the HSNJL model mechanism  \cite{034,042}, it is then plausible
to have a SSM with no input mass parameter for which soft supersymmetry
breaking and a subsequent electroweak symmetry breaking all being
generated dynamically within the model. The Higgs superfields are also
dynamical composites \cite{034,042}. All one has to do is to consider
higher dimensional operators of various four-superfield interaction terms
with some having strong couplings. A first soft breaking of supersymmetry as
illustrated here can generate the soft masses. Nonzero soft squark masses
together with appropriate holomorphic four superfield interaction(s) may
induce dynamical electroweak symmetry breaking and generates the masses
for the quarks, leptons, and gauge bosons. No other supersymmetry breaking
sector, messenger sector, or hidden sector is needed. It is a very simple
model without any hierarchy issue. All mass scales are generated dynamically.
Such a beautiful scenario of the origin of a phenomenological softly
broken SSM will be an interesting target for further studies.

Supersymmetry is a spacetime symmetry. Any theory with rigid supersymmetry
could be, and arguably should be, incorporated into a theory with
supergravity. In the latter case, the massless Goldstino will be eaten
up by the now massive gravitino. The model then should have special
implications for the couplings of the longitudinal part of the gravitino
to the matter superfields.

The central feature that distinguishes the model from other models of
supersymmetry breaking is of course the presence of the spin one
composite $v_\mu$ and its supersymmetric partners.

We take only the case of a simple singlet composite of
$U \sim \Phi^\dag_a \Phi^a$ here. A somewhat more complicated case as
studied in the case of (non-supersymmetric) NJL-type composite of spin one
field \cite{S} would have the composite in the adjoint representation.
Similar but superfield version of four-superfield interactions may be
considered though not in relation to pure soft supersymmetry breaking.
It is also possible to have a model in which the composite superfield
$U$ behaves like a massive gauge field supermultiplet \cite{064}, much
in parallel with the non-supersymmetric models of Ref.\cite{S}. It is
possible to think about the electroweak gauge bosons as such composites.
However, we echo the author of  Ref.\cite{S} against advocating the
kind of scenario.

Finally, we emphasize that with the modern effective (field) theory
perspective, it is the most natural thing to consider any theory as
an effective description of Nature only within a limited domain/scale.
Physics is arguably only about effective theories, as any theory
can only be verified experimentally up to a finite scale and there may
always be a cut-off beyond that. Having a cutoff scale with the so-called
nonrenormalizable higher dimensional operators is hence in no sense an
undesirable feature. Model content not admitting any other parameter
with mass dimension in the Lagrangian would be very natural. Dynamical
mass generation with symmetry breaking is then necessary to give the
usual kind of low energy phenomenology such as the Standard Model one.

The bottom line is, with relevancy for the supersymmetric standard
model or not, we present here a real simple model for dynamical
supersymmetry breaking, characterized
by the generation of soft mass(es) and a spin one composite.

\acknowledgments Y.C., Y.-M.D., and O.K. are partially supported
by research grant NSC 102-2112-M-008-007-MY3, and Y.C. further
supported by grants NSC 103-2811-M-008-018 of the MOST of Taiwan.
G.F. is supported by research grant NTU-ERP-102R7701 and partially
supported by research grants NSC 102-2112-M-008-007-MY3, and NSC
103-2811-M-008-018 of the MOST of Taiwan.

\end{document}